# Tuning the convergence angle for optimum STEM performance

**Matthew Weyland and David A. Muller**

Cornell University, School of Applied and Engineering Physics, Ithaca, NY 14853, USA

## ABSTRACT

The achievable instrumental performance of a scanning transmission electron microscope (STEM) is determined by the size and shape of the incident electron probe. The most important optical factor in achieving the optimum probe profile is the radius of the probe-forming aperture, which determines the convergence semi-angle of the illumination. What is often overlooked however is that small deviations from this optimum can degrade both the resolution and interpretability of image contrast. A 30% error in aperture radius can lead to a factor of 2 contrast reduction in typical lattice spacings, and a 5 Å error in the thickness measurement of thin layers (such as gate oxides). Theoretical calculations of the optimum convergence angles, from a wave-optical consideration of the probe forming conditions, are explained and their consequences discussed. An experimental approach to the measurement and tuning of the convergence angle is then introduced.

## Introduction

Scanning transmission electron microscopy (STEM), and in particular high angle annular dark field (HAADF) "Z-contrast" STEM, is becoming a key tool in the resolution of structural and functional problems at the atomic scale. The reasons for this are varied but the most compelling are the combination of the high resolution, on the order of an Angstrom, and the interpretability of the image contrast, which is strongly sensitive to the atomic number of the scattering atoms and the mass-thickness of the specimen[1-4]. While there are a large number of factors that control the ultimate performance of a STEM instrument, including electron source brightness, accelerating voltage and lens aberrations, many of these are set by the design and specification of the microscope. The operator is left to optimize the gun settings and lens strengths and choose the probe-forming aperture that determines the convergence semi-angle ($\alpha$) of the electron probe. This aperture is known as the objective aperture in dedicated STEM instruments and the condenser aperture in a combined TEM/STEM (to avoid confusion the term probe forming aperture is used throughout this paper). With the convergence angle fixed, source size can be traded off against probe current.

The choice of probe forming aperture, and the resultant $\alpha$, is often overlooked (with devastating consequences) as the choice is often compared to the choice of objective aperture in parallel beam illumination. The selection of an objective (post specimen) aperture in TEM is usually made to improve contrast in the recorded image through exclusion of scattered electrons. High resolution (lattice images) in Bright-field TEM are achieved as long as the aperture is not too small for the collection of electrons scattered to the Bragg spot associated with the lattice spacing required. Indeed even without an objective aperture a high resolution TEM image can be obtained by allowing the entire range of frequencies to be included, damped by the natural contrast transfer function (CTF) of the instrument. Essentially the objective aperture is a linear filter determining the frequencies to be included in the TEM image, for example in a weak phase object filtering out the high frequency oscillations in the CTF which show inverted contrast, see Fig.1 a). Increasing the aperture size does not affect the lower frequencies. However the effect of aperture radius on the CTF for HAADF STEM imaging is a far more complex issue. The ADF image point spread function is proportional to the square of the probe wavefunction and not the wavefunction itself (as in Bright field TEM). The resultant response to increasing the aperture size is non-linear and affects all spatial frequencies, not just the highest (squaring the wavefunction in real space is equivalent to self-convolving its Fourier transform in diffraction space, so high and low frequencies are mixed). Essentially phase errors at large angles to the optic axis are mixed in to the lower frequencies (which would be unaffected in BF), degrading both contrast and image localization, even as the information limit is increased.

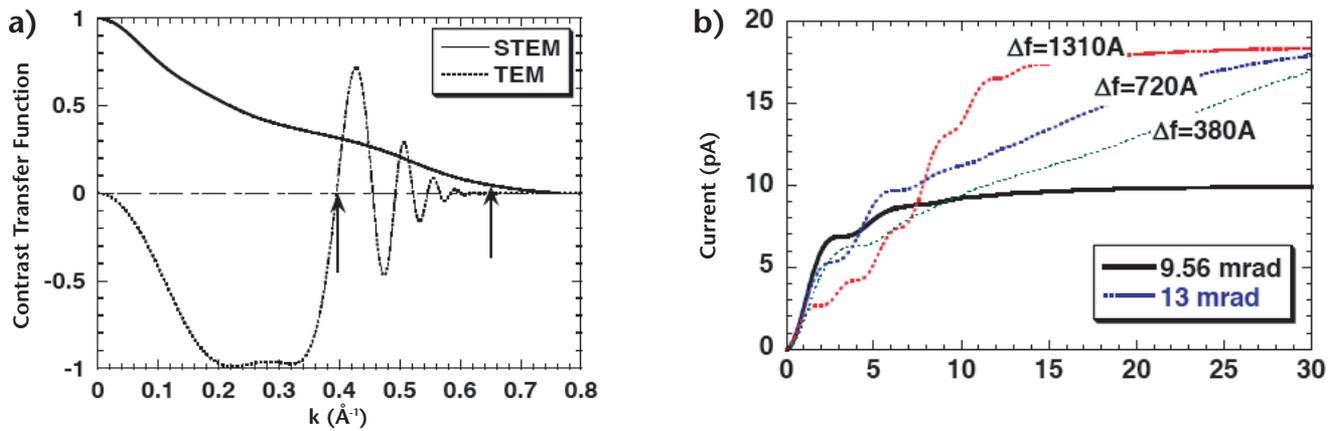

Figure 1. a) Contrast transfer function (CTF) for a conventional, almost-parallel beam, TEM calculated for an FEI tecnai F20 SuperTWIN. The dashed arrow marks the potential position of the objective aperture "cutoff" which would exclude contrast reversals from higher frequencies in the resultant TEM image. The STEM CTF is calculated for this sized aperture, has a 5% information limit almost double the aperture size (solid arrow). b) Beam current enclosed within a given diameter for the F20-ST (200 kV, Cs=1.2 mm, 1Å source size) for the optimal 9.6 mrad and too-large 13 mrad aperture as an illustration of the effect of condenser aperture size on STEM 2D probe point spread function (PSF). The larger aperture provides almost double the beam current, but all of this extra current falls outside the central peak. Increasing the aperture size beyond the optimum only reduces the signal/background ratio.

A holistic approach to optimizing the convergence semi-angle ($\alpha$) is described, based on the wave optical theory of probe formation. The experimental method for calibrating the convergence angles of probe forming apertures in the STEM is described and suggestions made on how to fine tune any mismatch between theoretical optimum and instrumental values.

## Calculating optimum convergence angles

There are three classical contributors to the probe shape in the electron microscope; the effect of a finite source size (the 'gun') contribution, the aberrations induced by the primary imaging lens (the spherical aberration ($C_s$) term), and the diffraction limit. In a simple geometric optics approach these factors can be assumed to be Gaussian and hence can be added in quadrature[5, 6]. Achieving optimum performance is a balance between $C_s$ and diffraction, see Fig. 2. The source size term due to the finite gun brightness can also be added in quadrature and has the same angular dependence as the diffraction limit. While this approach is suitable for a large analytical probes where the source size term can be dominant, it consistently both overestimates the probe size and underestimates the optimum convergence semi-angle for high-resolution STEM imaging. A more accurate wave-optical formulation[7, 8] can be applied based on the Scherzer aberration function[9] which describes the phase shift of a wave at angle α to the optic axis, $\chi(\alpha)$:

$$\chi(\alpha) = \frac{2\pi}{\lambda}\left(\frac{1}{4}C_s\alpha^4 - \frac{1}{2}\Delta f \alpha^2\right) \qquad -(1)$$

where $\lambda$ is the electron wavelength and $\Delta f$ is the defocus value. Spherical aberration from a round lens causes the rays off-axis to be deflected more strongly than for an ideal lens, leading to a positive phase shift. This can be partially compensated for a limited band of spatial frequencies by applying a negative defocus (Fig. 3). An ideal lens for ADF has zero phase shift across the aperture. In practice, a maximum allowable phase shift of $\pi/2$ (i.e. a quarter wavelength) across the objective lens is tolerated and solving for defocus and maximum angle, $\alpha_0$ allows the assessment of the optimum probe forming conditions. A detailed derivation is given in Appendix A. This leads to a simple expression for both the minimum ($d_0$) attainable full-width half maximum (FWHM) probe size and the optimum convergence semi-angle ($\alpha_0$) [9]:

$$d_0 = 0.43\, C_s^{1/4}\lambda^{3/4}\,, \qquad \alpha_0 = \left(\frac{4\lambda}{C_s}\right)^{1/4} \qquad -(2)$$

If these are calculated for a 200 kV FEI Tecnai F20 SuperTWIN ($C_s$=1.2mm) the minimum probe size achievable is 1.6 Å with an optimum convergence semi-angle of 9.6 mrad. This is a significantly higher performance than that suggested by the non wave-optical methods (~2.8Å for an ~7 mrad aperture). The aberration function can then be integrated numerically over the probe-forming aperture and squared to generate the point spread function, and if Fourier transformed the contrast transfer function (CTF), for particular α and $\Delta f$ [7, 8].

Example plots of PSF and CTF are shown for optimum (9.6 mrad) and spherical aberration limited (13 mrad) conditions in Fig. 4. The optimum defocus for 9.6 mrad ($\Delta f_{opt}$ ~ 500 Å) generates a probe with a simple CTF, Fig. 5, and minimal probe tails in the PSF. The contrast with the optimal aperture is roughly 2-3 times larger at low spatial frequencies than for the bigger aperture. There is also only 1 defocus setting at which the image will look the "sharpest". However for the 13 mrad aperture, there are 2 defocus settings which are local maxima. Focusing by eye will more likely lead to focusing at the secondary optimal defocus ($\Delta f_{opt2}$ at ~ 1300Å), which is due to the strong peaks in the CTF at typical lattice spacings at this defocus value, see Fig. 6. At this defocus however the accumulated phase shifts across the aperture also causing substantial delocalization and large probe tails, see Fig. 4 d). This has the effect of increasing the FWHM of the probe and causing the probe current to be spread outside the central maximal reducing the contrast from an imaged lattice and cause a large delocalization of any analytical signal (such as EELS) recorded. A further problem is the lack of a balanced frequency response with the larger aperture. While high frequencies are attainable with the 13 mrad convergence at higher defocus values, they come at the expense of a loss in response for lower frequencies. Effectively, the contrast of larger lattice spacings will decrease, even disappear entirely, as the smaller spacings come into focus.

The imaging problems associated with a spherical aberration limited probe are examined in Fig 7 for a Si/SrTiO$_3$ interface. While both images are acquired with non optimal convergence semi-angle, the spherical aberration limited example shows significant problems. The large probe tails spread the intensity of the high-Z SrTiO$_3$ layer into the Si resulting in a slope of increased intensity towards the interface. The width of the interface itself also appears significantly broader and there is a significant loss in contrast from the atomic columns in both layers.

If a very thin, low-Z layer is sandwiched between two high-Z layers (such as an accidental SiO$_2$ layer between a HfO$_2$ gate oxide and its silicon substrate), the probe tails from an oversized aperture can wash out the details of the light layer completely, and sometimes it will not even be detectable above the background. For slightly thicker low-Z layers (10-20 Å), tails from each of the heavier layers can still overlap and attempts to measure the thickness of the low-Z layer will overestimate its width. This can lead to errors as large as 5 Å in measuring the width of 15 Å thick gate oxide – more details of this metrology problem are covered in the article by Diebold et al. [4].

Although using an oversized aperture is clearly a disaster for analytical work – the delocalized tails make atomic resolution analysis impossible. The increased information limit makes it useful for high-resolution imaging of perfect crystals and testing the information limit of a STEM [1, 10].

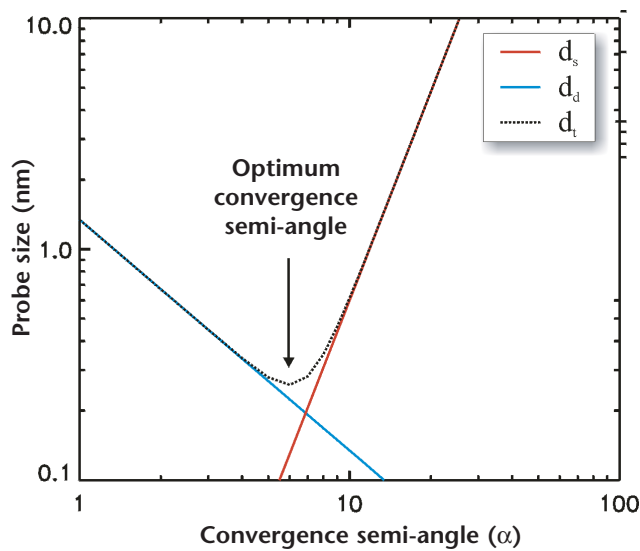

Figure 2. Balancing spherical aberration against diffraction. At low convergence semi-angle ($\alpha$) the diffraction contribution ($d_d$) term dominates, while at high $\alpha$ the spherical aberration contribution ($d_s$) is dominant. The terms are calculated by $d_d = 0.61\lambda/\alpha$ and $d_s = {}^1/_2 C_s \alpha 3$. Terms are added in quadrature to generate the total ($d_t$). Note resultant convergence angle is 6-7 mrad, to give a probe size of ~2.8Å, which is a more pessimistic estimate than the wave optical treatment.

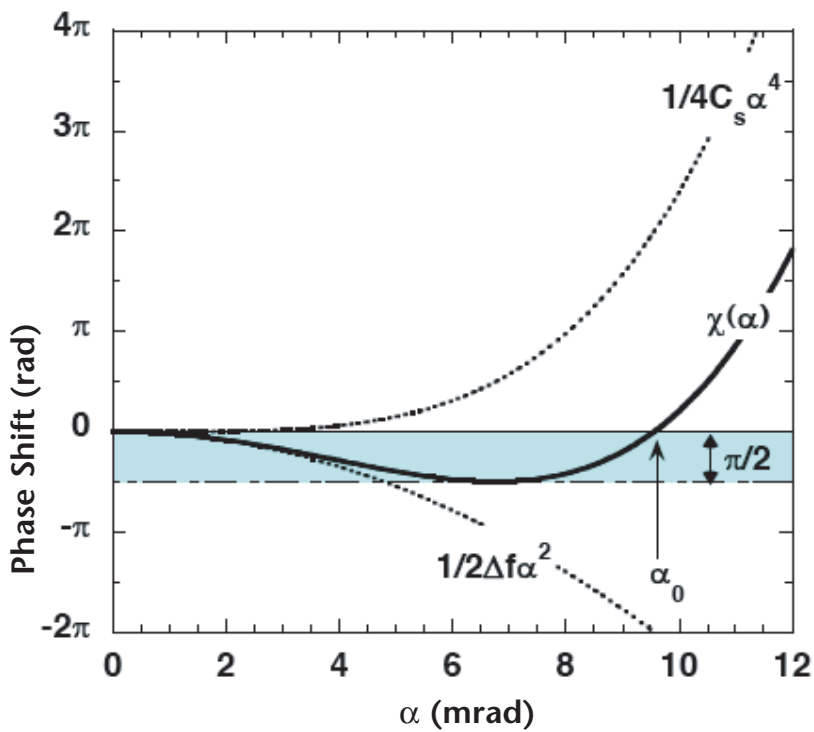

Figure 3. Balancing spherical aberration against defocus to obtain a uniform phase shift across the probe forming aperture. At low convergence semi-angle ($\alpha$) a negative defocus contribution (the $1/2\Delta f \alpha^2$ term) can cancel the spherical aberration contribution ($1/4C_s\alpha^2$ term), while at high $\alpha$ the spherical aberration dominates the aberration function $\chi(\alpha)$. An ideal lens would have zero phase shift, but a tolerable error is usually considered to be a quarter wavelength, i.e. a $\pi/2$ band shown by the shading, which sets both the optimal defocus and the maximum aperture size, $\alpha_0$.

## Ronchigrams, selecting apertures and measuring convergence semi-angles

The ideal approach to selection and alignment of the correct aperture is to make use of the coherent Ronchigram formed on an area of amorphous material[11, 12] (using the largest probe-forming aperture in order to see all the details in the Ronchigram). The Ronchigram is the convergent beam diffraction pattern of an amorphous region or a crystal where the probe forming aperture is much larger than the Bragg angles. Out of focus, the Ronchigram gives a shadow image of the sample in the diffraction plane. In the absence of lens aberrations, if the beam is focused before the sample (Fig. 8a), the Ronchigram is an erect, magnified image of the illuminated portion of the sample (and the magnification is the camera length/defocus). When the beam is focused after the sample (Fig. 8b), the Ronchigram is inverted. When the beam is at crossover on the sample (Fig. 8c), the Ronchigram is an image at infinite magnification (defocus is 0), which should look smooth and featureless for a very thin sample.

When the lens has spherical aberrations, the beam can only be brought to crossover for small angles (Fig. 8d). At large angles, the beam must cross before the sample, leaving a distorted shadowed image to surround the small disk of infinite magnification. Image reversals from inverted to erect must occur for all overfocus settings, leading to rings in the Ronchigram (if the probe is properly stigmated). This focused Ronchigram, Fig. 9, is a reflection of the imaging optics, with the "rings" surrounding the central region of the interference pattern a consequence of $C_S$ and the smooth region in the center is an image of the amorphous specimen at infinite magnification, with the frequency of the image information increasing with the radius. (If the sample is more than a few nm thick, the smooth region will be replaced by faint random mist, as the top and bottom cannot both be at infinite magnification at the same defocus). Where the rings start indicates the frequency at which $C_S$ dominates and an aperture should be chosen that is small enough not to include these rings. However too small an aperture will not include the highest frequencies available, and the probe becomes diffraction limited. An ideal aperture will sit just inside the rings. Choosing, and aligning, the aperture is best achieved by marking the center of the ronchigram (either by the beam stop or mark on the small screen) and changing apertures until one is found of the correct size. With the correct aperture it becomes difficult to center the aperture on the Ronchigram (as the small size hides much of the Ronchigram detail), but the aperture can be centered with reference to the beam stop/mark. Ideally the condenser and objective optics of the microscope should be such that one of the probe forming apertures is almost precisely the correct size to give the optimum convergence semi-angle. However in reality this is rarely the case and to achieve optimal performance the balance of the probe forming lenses will have to be adjusted.

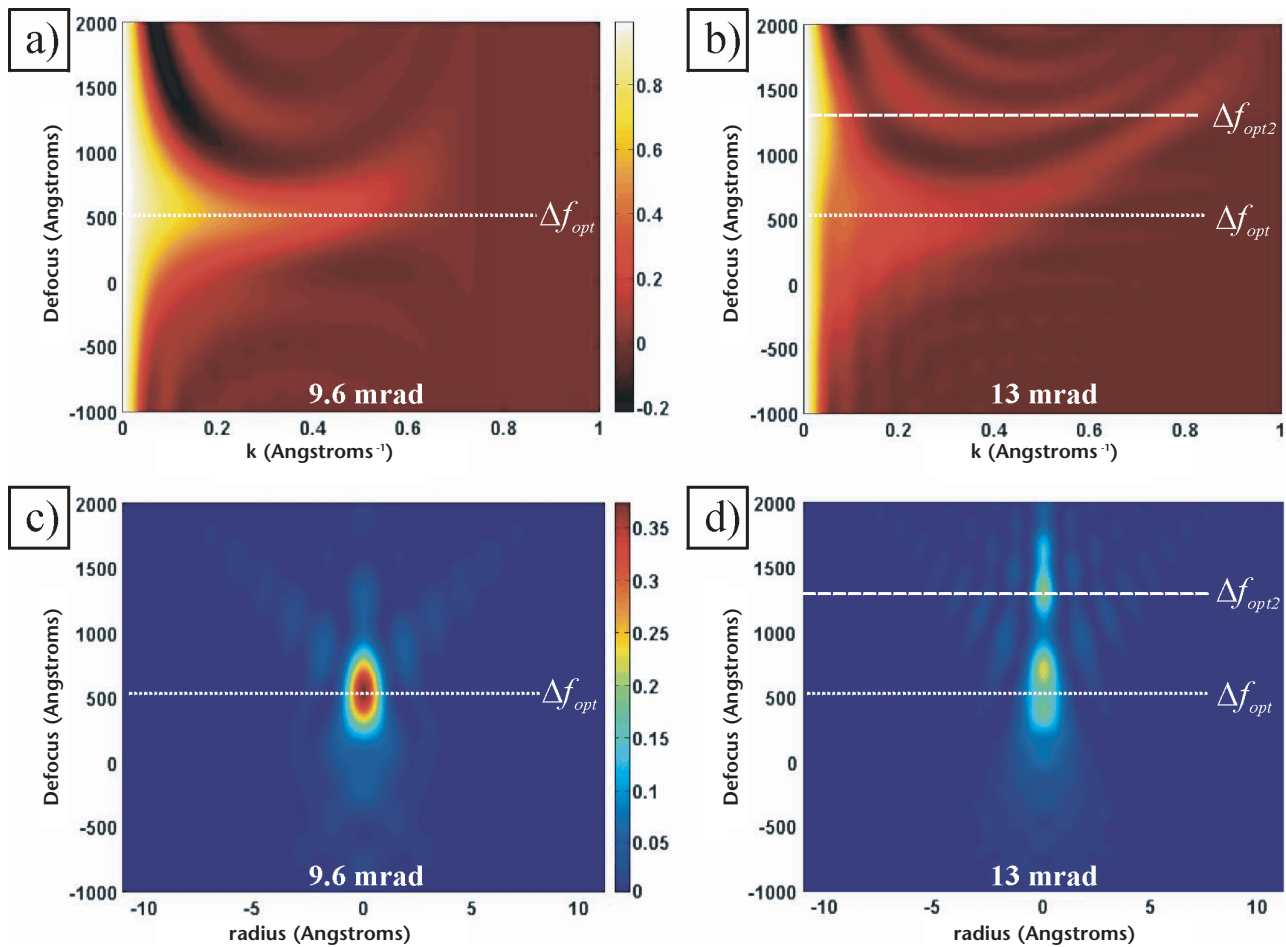

**Figure 4.** Contrast transfer functions (CTF) and point spread functions (PSF) calculated from wave optical considerations. All calculations carried out for the equivalent of a tecnai F20 SuperTWIN: 200kV accelerating voltage, $C_s$ of 1.2mm and a defocus range -1000 to +2000 Å. a) PSF and c) CTF for a 9.6 mrad convergence angle. b) PSF and d) CTF for a 13 mrad convergence angle. Marked are optimal defocus settings ($\Delta f_{opt}$), for a larger aperture there is a second optimal ($\Delta f_{opt2}$) due to the secondary maxima in defocus.

The experimental method for measurement of the convergence semi-angle is fairly simple; relying on the calibration of the CBED pattern in the diffraction (detection) plane of the STEM. In a combined TEM/STEM instrument this is usually conjugate with the viewing screen, which can be directly captured on a CCD camera (or plate film if this is not available). In a dedicated STEM instrument the diffraction pattern from a stationary probe can be formed on the BF detector by rastering the beam through reciprocal space using the post specimen scan coils.

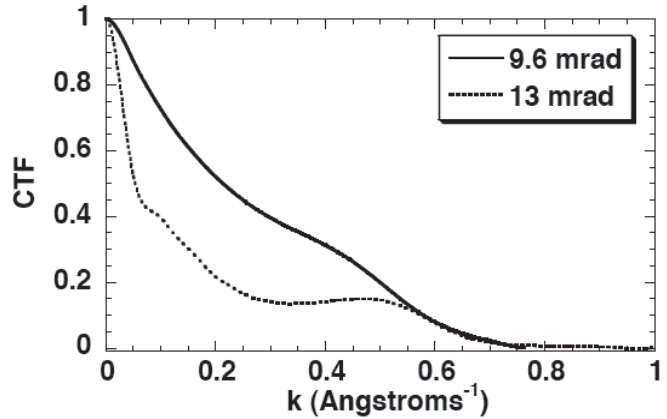

Figure 5. Contrast Transfer Function for the T20-ST (200 kV, Cs=1.2 mm, 1Å source size) for the optimal 9.6 mrad and too-large 13 mrad aperture at the defocus setting to give the most-peaked point-spread function for each aperture. Notice how the larger aperture reduces the contrast at lower spatial frequencies two to threefold.

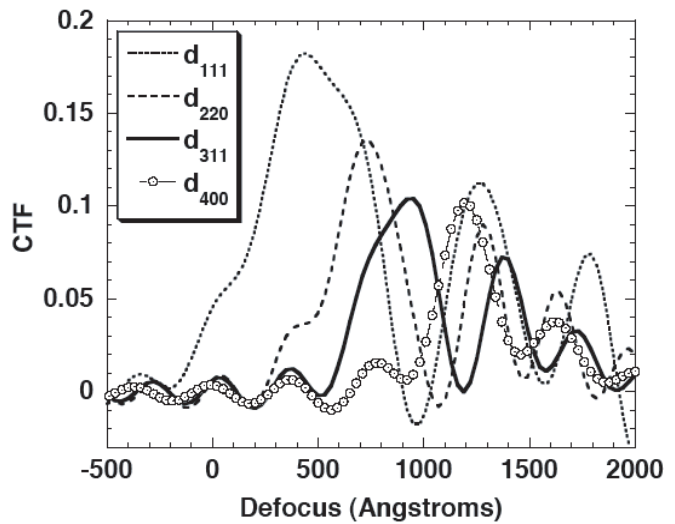

Figure 6. Contrast Transfer Function for the T20-SuperTWIN (200 kV, Cs=1.2 mm, 1Å source size) as a function of defocus for the first 4 lattice spacings of silicon (3.13, 1.92, 1.63, 1.36 Å respectively). Due to the phase shift across the aperture, higher spatial frequencies are transmitted through the lens most efficiently at different defocus settings from the lower spatial frequencies. A user focusing by eye (or FFT) for the sharpest image will likely pick the secondary maximum at 1300 Å which contains all spatial frequencies.

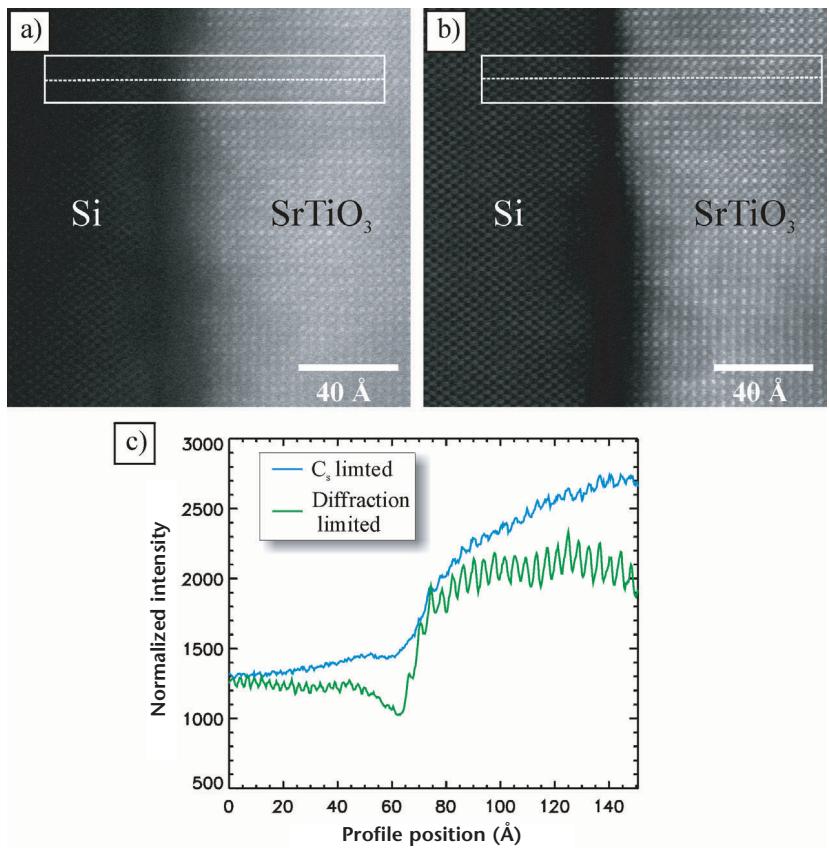

Figure 7. Effect of aperture radius/convergence angles on apparent interface width between Si and SrTiO$_3$. a), b) STEM HAADF images acquired from the same area with a) $C_s$ and b) diffraction limited convergence semi-angle respectively. c) Line traces (averaged and normalized) through the interface, at the positions marked, for the two conditions. These profiles clearly show the effects of the probe tails: a "spreading" of the intensity of the SrTiO$_3$ layer into the Si, apparent broadening of the interface and the drop in contrast on the atomic columns.

With a known crystal and a known orientation the ratio between the spacing between Bragg discs ($b$) and the aperture diameter ($a$) will be proportional to the Bragg angle ($\theta_b$) for a particular reflection (for a given wavelength) and the convergence semi-angle ($\alpha$) of the probe:

$$\frac{a}{b} = \frac{\alpha}{\theta_b}$$

By measuring $a$ and $b$ for a known spacing allows $\alpha$ to be determined. A common reflection chosen for this calculation is silicon (200) or (220), as demonstrated in Fig 10. While there are usually a handful of different probe-forming apertures in a given instrument the spacing between the Bragg discs will not vary between them (as this is determined by the Bragg angle and the camera length): as such the full pattern need only be recorded once as the rest of the convergence angles may be measured by merely measuring the diameter of the central Bragg disc ($a$) for each aperture (provided the lenses and the sample are not adjusted). In Fig. 10 the experimental measurements of convergence semi-angle for the four probe forming apertures is: 5.6, 8.0, 11.1 and 16.9 mrad. None of these apertures is close to the 9.6 mrad determined as the optimum semi-angle from the wave-optical calculations.

## Tuning convergence semi-angle

While the choice of aperture has a large influence on the convergence semi-angle it is not the only factor involved: the imaging optics, in particular the C2 and objective lenses, have significant influence over $\alpha$ and are easier to modify than the diameter of the aperture. In STEM imaging the final condenser lens (C2) can be used to focus, in the plane of the specimen and probe, in much the same way as the objective lens. Indeed in modern TEM/STEM instruments it is usual to fix the objective lens at the optimum value for eucentric focus and adjust C2 to focus. However the angles subtended by both lenses are different and carrying out an equivalent defocus with both lenses will not result in an equivalent change in convergence semi-angle. As such it is possible, by modifying one and compensating with the other, to stay in focus yet change the effective α. A series of semi-angles have been measured for varying excitations of the objective lens for a Tecnai F20 SuperTWIN, Fig.11 a). It is apparent that the objective lens range suitable for forming a small STEM probe is very limited, with a 3% percent change in objective lens strength (and the balancing change in C2 to bring to focus) covering a convergence angle change of approximately 7 mrad (~4-11mrad). This illustrates the importance of the eucentric focus setting in a modern TEM/STEM, providing a fixed value of objective lens strength at a certain specimen height removes the free variables that would make achieving acceptable STEM performance in a combined system impractical.

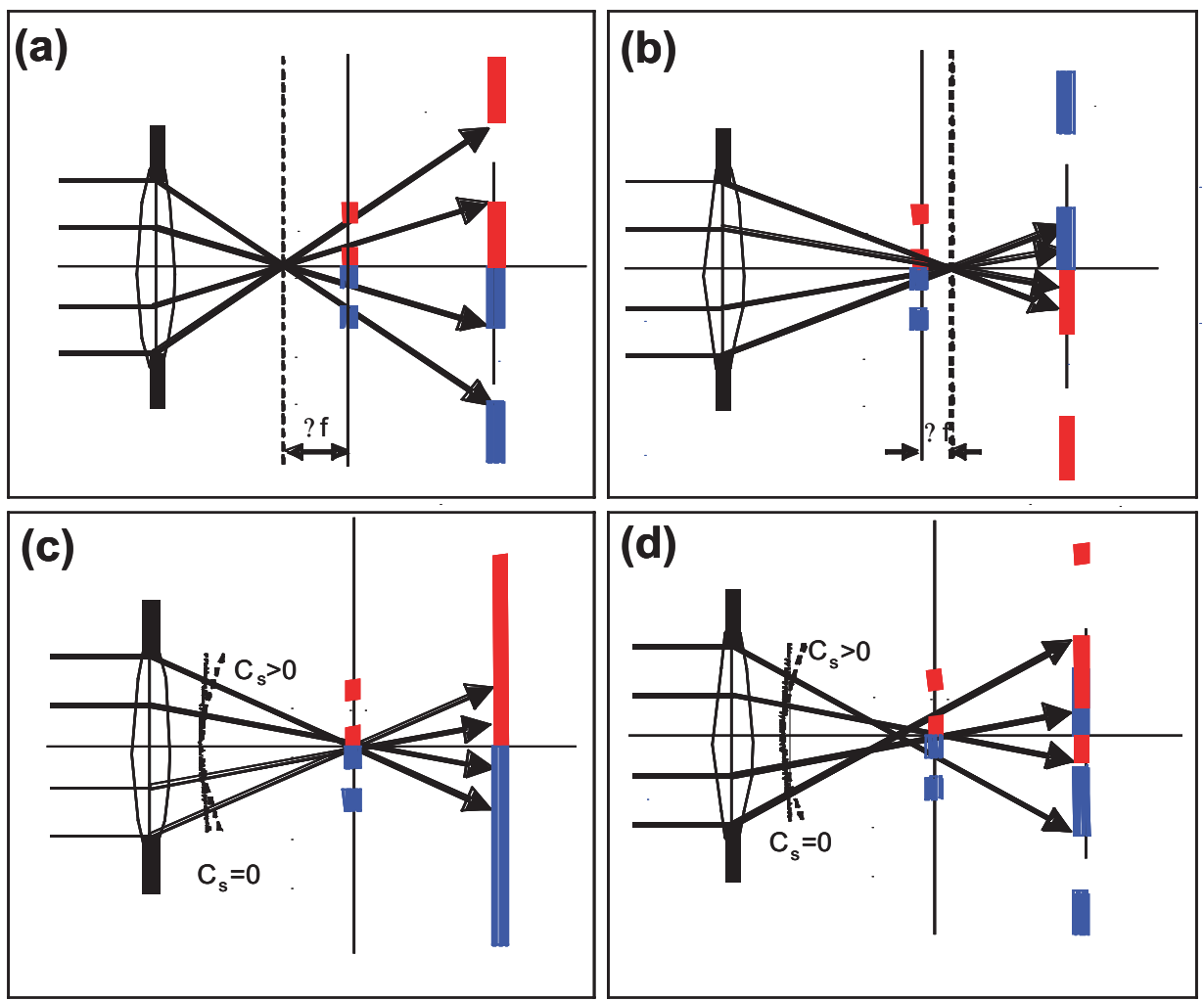

**Figure 8.** Ray diagrams showing the formation of Ronchigram shadow images for different defocus conditions. Cases (a)-(c) are in the absence of spherical aberration. (a) If the probe is focused to a crossover before the sample, the shadow image projected on to the viewing screen is magnified and erect, with the magnification being the ratio of the camera length/defocus. (b) If the crossover is after the sample, the shadow image is magnified and inverted. (c) In the absence of lens aberrations, if the beam is near cross over, the image magnification is almost infinite. (d) Near cross-over with spherical aberration - rays off-axis to come to a focus before the sample. The region of near-infinite magnification can only be maintained at small angles.

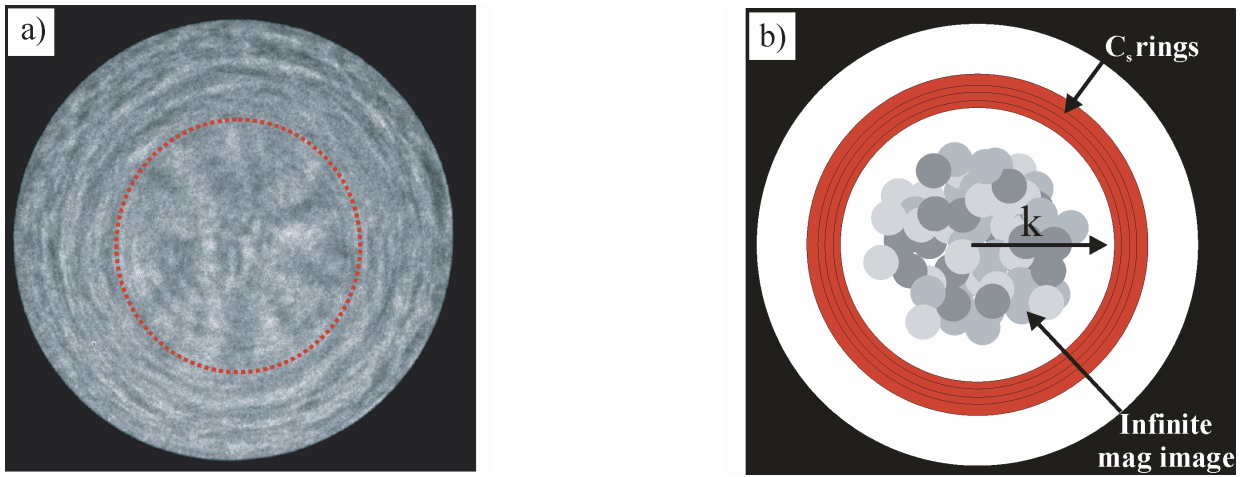

**Figure 9.** Description of the coherent electron ronchigram (formed from amorphous material). a) An experimental ronchigram, formed inside the largest probe forming aperture. b) A simplified conceptual "map" of the ronchigram showing the location of the rings caused by $C_s$, and the presence of the shadow image inside the rings. In the real case the "blobs" in the center of the ronchigram flash randomly due to the small movements of the incident probe and specimen. The spatial frequency (k) of the STEM image will increase with the radius of the Ronchigram selected out by the probe-forming aperture, however too large an aperture will include the $C_s$ rings. Therefore the ideal location for the probe-forming aperture (which defines the convergence semi-angle) is just inside the $C_s$ rings and is marked with a dotted line in the experimental Ronchigram.

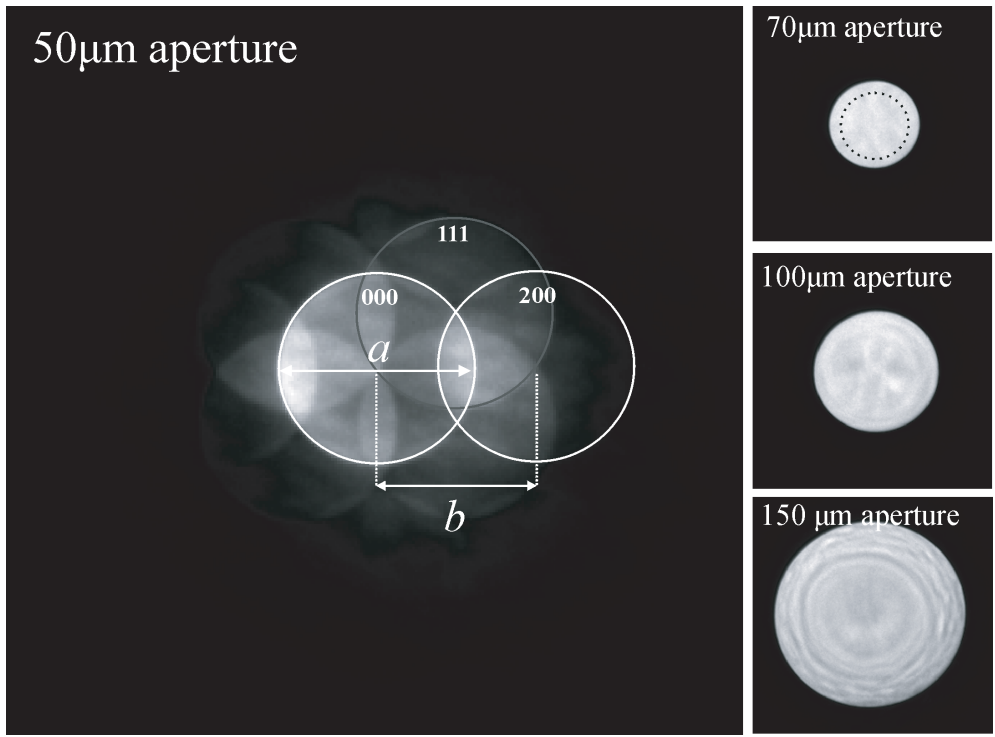

**Figure 10.** Measurement of STEM convergence angles in an FEI F20 SuperTWIN. The diffraction pattern is "calibrated" on the the 200 reflection of Si, oriented onto to the 110 axis. The convergence semi-angle ($\alpha$) is proportional to the ratio of the disc width to the disc spacing ($a/b$). As $b$ is independent of the chosen aperture the other three apertures can be calibrated by just recording the width of the zero order disk ($a$). The dotted line inside the 50μm aperture represents the relative scale of the 50μm aperture.

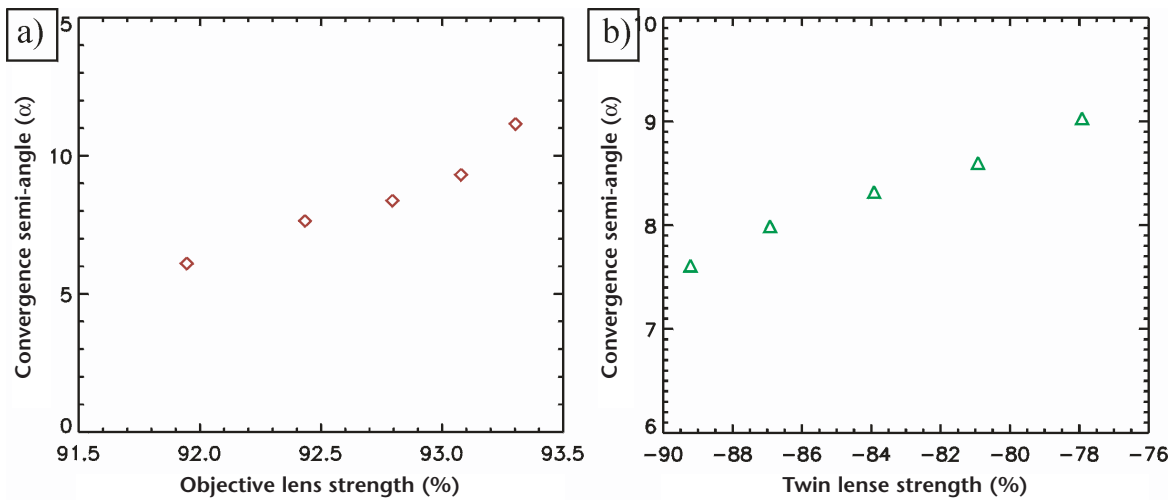

Figure 11. Effect of lens strengths on measured convergence semi-angles. For a) Objective lens and b) Twin (mini-condenser) lens values. In all cases the specimen was refocused using C2. Note the twin lens values are –ve due to the convention in FEI instruments, a –ve value indicates a converged beam mode (STEM) while +ve is for parallel beam (CTEM).

A short through-convergence series, of the kind carried out here, should be sufficient to determine a more accurate balance of objective/C2 to achieve an optimum convergence semi-angle. An alternative approach is to adjust the minicondenser lens (the TWIN lens on FEI systems), balancing out with C2 or objective. The minicondenser sits between the C2 and objective lenses (on FEI systems it is actually part of the objective lens) and is instrumental in switching between parallel and convergent (STEM) illumination. Changing this lens value, and balancing with C2, also has a clear effect on the semi-angle, Fig. 11 b).

With appropriate lens settings almost any of the physical apertures can be tuned to the optimal convergence angle. For a given convergence angle, the larger apertures will produce a larger source size blurring and proportionally larger beam current (because beam brightness is fixed). This again can be compensated by the gun lens, so the final choice of apertures will depend on where the microscope is most sensitive to stray fields and external noise.

## Conclusions

Calculations of point spread and contrast transfer functions from wave-optical considerations clearly show that to combine optimum resolution with interpretable image contrast requires tight control over convergence semi-angle. The approach to measurement of the experimental semi-angle has been described and this allows, in combination with calibrated adjustments to the probe forming lenses, the optimization of microscope performance. This approach should lead to more certain measurement of features such as interface layers as well as optimal interpretable contrast in the lattice image.

## Appendix A – Deriving the Optimal Aperture Size for STEM Imaging

Scherzer's 1949 paper on spatial resolution in electron microscopy contains the now often-cited wave optical estimate of an optimal aperture size for incoherent imaging in the presence of spherical aberration [9].

A full derivation is given here so the reader can understand both the underlying assumptions and how to generalize the approach to include fifth order aberrations (which involves balancing Cs and defocus against the higher order terms to keep the phase errors small).

We start by writing the electron wavefunction in the front focal plane as a function of scattering semi-angle $\alpha$ as

$$\varphi(\alpha) = e^{i\chi(\alpha)} \quad \text{-(A.1)}.$$

where the phase introduced by the lens is

$$\chi(\alpha) = \frac{2\pi}{\lambda}\left(\frac{1}{4}C_s\alpha^4 - \frac{1}{2}\Delta f\alpha^2\right) \quad \text{-(A.2)}.$$

The shape, or rather, the point spread function (PSF) of the probe can be calculated by Fourier transforming and squaring the wavefunction. $d_0 = 0.61\lambda/\alpha_0$ For an ideal lens, $\chi(\alpha) = 0$, and finite aperture $\chi(\alpha)_0$, the PSF becomes the square of the Airy function with width, $d_0 = 0.61\lambda/\alpha_0$, in keeping with Lord Rayleigh's resolution criterion[13].

Our goal here is to find the maximum aperture size, $\alpha_0$, where the phase shift across the aperture is kept tolerably small by balancing $\Delta f$ against $C_s$. Scherzer chose to allow a maximum phase error of a quarter wavelength – i.e. $|\chi(\alpha)| \leq \pi/2$ for all $\alpha \leq \alpha_0$ If the maximum tolerable phase error is changed from $\pi/2$ only the numerical prefactors in equation 2 (main text) are altered.

The problem can be simplified by writing so $x = \alpha^2$

$$\chi(x) = \frac{\pi}{\lambda}\left(\frac{C_s}{2}x^2 - \Delta f x\right) \quad \text{-(A.3)}.$$

The minimum of $\chi(x)$ occurs at

$$\left.\frac{\partial \chi(x)}{\partial x}\right|_{x=x_{min}} = \frac{\pi}{\lambda}(C_s x - \Delta f) = 0 \quad \text{-(A.4)}.$$

i.e.

$$x_{min} = \frac{\Delta f}{C_s} \quad \text{-(A.5)}.$$

The optimal defocus can now be found by noting that we want $\chi(x_{min}) = -\pi/2$ Substituting (A.5) into (A.3) we find

$$\Delta f_{opt} = (C_s\lambda)^{1/2} \quad \text{-(A.6)}.$$

The largest aperture size $\alpha_0$, is set by the point where $\chi(\alpha) = 0, \alpha > 0$ in Fig. 3. Setting (A.3) equal to 0, we find

$$\left(\frac{C_s}{2}x - \Delta f\right)x = 0 \quad \text{-(A.7)}.$$

For $x \neq 0$ we get the optimal $x_0$ as a function of defocus

$$x_0 = \frac{2\Delta f_{opt}}{C_s} \quad \text{-(A.8)}.$$

We now use equation (A.6) to eliminate the optimal defocus, and note that $\alpha_0 = x_0^{1/2}$ to get

$$\alpha_0 = \left(\frac{4\lambda}{C_s}\right)^{1/4} = 1.41\left(\frac{\lambda}{C_s}\right)^{1/4} \quad \text{-(A.9)}.$$

Note that this optimal aperture size is smaller than the optimal aperture size for coherent imaging in TEM ($1.56\,(\lambda/C_s)^{1/4}$), which is also called the Scherzer aperture.

The basic assumption at the start of this derivation was that all the phase shifts at angles less than $\alpha_0$ were required to be small so the image blur, $d_0$, will be limited by the diffraction limit for incoherent imaging with aperture size $\alpha_0$,

$$d_0 = \frac{0.61\lambda}{\alpha_0} = \frac{0.61}{\sqrt{2}}C_s^{1/4}\lambda^{3/4} = 0.43\,C_s^{1/4}\lambda^{3/4} \quad \text{-(A.10)}.$$

Equations (A.9) and (A.10) are the desired result for equation (2) in the main body of the text.

It is interesting to note that in light optics, the tolerable phase error is usually taken to be a tenth of a wavelength or less, i.e. solving for $\chi(x_{min}) = -2\pi/10$ With these constraints we find instead,

$$d_0 = 0.54\, C_s^{1/4} \lambda^{3/4}, \qquad \alpha_0 = 1.124 \left(\frac{\lambda}{C_s}\right)^{1/4} \quad \text{-(A.11)}.$$

Reducing the phase error from $\lambda/4$ to $\lambda/10$ togives a 20% worse spatial resolution and a 40% reduction in beam current (which is proportional to $\alpha_0^2$). Kirkland has calculated the point spread function that minimizes the probe tails, but at the price of increasing the central disk[7]. This gives $\alpha_0 = 1.22(\lambda/C_s)^{1/4}$ which falls between our two estimates. Colliex and Mory have also considered wave optical estimates of probe size as a function of aperture size and defocus [8]. The precise value of that is chosen depends on the compromise one is willing to make between achieving the smallest probe full width at half maximum (large $\alpha_0$) and the smallest probe tails (smallest phase shift error).